\documentclass[epj,final]{svjour}
\usepackage{graphics,epsfig,rotating}
\begin{document}
\title{Limits of complete equilibration of fragments produced in central Au on Au collisions at intermediate energies} 

\author{
     W. Neubert\inst{1}
\and A.S.Botvina\inst{2,3}
}
\institute{Institut f\"ur Kern- und Hadronenphysik, Forschungszentrum Rossendorf, 
     01314 Dresden, Germany  
\and Gesellschaft f\"ur Schwerionenforschung, 64291 Darmstadt, Germany
\and Institute for Nuclear Research, Russian Academy of Science,
     117312 Moscow, Russia  
}
\date{Received: \today}
\abstract{ 
Experimental data related to fragment production in central Au on Au 
collisions were analysed 
in the framework of a modified statistical model which considers 
cluster production both prior and at the equilibrated stage. 
The analysis provides limits to the number of nucleons and to the 
temperature of the equilibrated source. The rather moderate temperatures 
obtained from experimental double-yield ratios of $d$,$t$,$^3$He and $^4$He are 
in agreement with the model calculations. 
A phenomenological relation was established between the collective 
flow and the chemical temperature in these reactions. 
It is shown that dynamical mechanisms of fragment production, e.g.  
coalescence, dominate at high energies. It is demonstrated that coalescence may 
be consistent with chemical 
equilibrium between the produced fragments. The different meaning of 
chemical and kinetic temperatures is discussed.  
} 


\maketitle

\section{Introduction}

The mechanism of fragment production at intermediate energy nucleus-nucleus 
collisions is a long standing problem which is important for many investigations, 
such as the study of the nuclear liquid-gas phase transition, the behaviour 
of nuclear matter under extreme conditions, and, in general, the 
nuclear equation of state. In this paper we concentrate on central collisions 
which deposite a large amount of energy into nuclei, and produce a fast 
explosion of 
nuclei into many fragments. Presently, there is evidence that at low 
projectile energies of $\sim$10--100 A$\cdot$MeV the fragment production via 
multifragmentation of thermal-like sources is the dominating process 
\cite{Bon95,agos,Lav,Bouriq}. With increasing energy
dynamical effects, such as collective flow, becomes prominent. 
This implies that the fragmentation mechanism changes
from a statistical to a dynamical one.
However, the description of intermediate mass fragment (IMF) production
as a result of the break-up of
an equilibrated source with collective flow is very 
successful, as shown in many publications \cite{Lav,will,neu2}.
We believe that such good descriptions were possible since the 
hypothesis of chemical equilibrium between different kinds of
fragments is adequate for these reactions. 

In this paper we analyse data from 100 to 1000 A$\cdot$MeV. 
Some of these data were already analysed with different dynamical and 
statistical approaches. The present analysis involves new degrees of freedom. 
In particular, it emphasizes the importance of isospin characteristics 
of produced fragments for the determination of the reaction mechanism. 

 As suggested by many theoretical and experimental studies 
\cite{Lav,will,bon2,Des,heid,Schar,hxi} 
the fragments may be produced in 
a fast initial ('preequilibrium') process as well as at the full
equilibration stage. 
Since the number of nucleons 
participating in the thermal-like source can decrease rapidly with the beam 
energy, dynamical processes of fragment formation should contribute 
essentially at higher energies. The mechanism of 
coalescence of nucleons into fragments is applied here for a complementary 
description of light fragment production to the statistical approach. In the
following we discuss a relation between coalescence and statistical 
approaches and we point out that the coalescence mechanism may simulate 
the chemical equilibrium conditions. In this respect, we pay 
special attention to the chemical temperature and study its correlation 
with the light charged particles (LCP) velocities. 
However, our main aim is to determine properties of a completely
equilibrated source, which is mainly responsible for IMF production. 
We emphasize, that the conclusions about equilibrium and nonequilibrium 
contributions concerns the production of fragments from nucleons only. We 
do not consider the problem of the nucleon thermalization and refer to 
experimentally selected central events which belong to an ensemble of nucleons
in some degree of equilibration. 
The knowledge of the relation between equilibrium and non-equilibrium 
mechanisms of fragment production in dynamical processes is important 
for many applications both in nuclear physics and astrophysics. For example, 
similar processes are expected during the fast synthesis of elements in the
early universe and in supernova explosions \cite{wallerstein}.

\section{Chemical temperatures evaluated from fragment data obtained in central
 Au+Au collisions}

Data obtained by the FOPI collaboration 
at 100, 150, 250 and 400 A$\cdot$MeV \cite{reis1,pogg} and data from the EOS
collaboration \cite{eos,lisa,scott} at 1 A$\cdot$GeV are analysed. Central events were
selected by the criterion ERAT as described in ref.~\cite{reis1} whereas the 
event centrality of the 1 A$\cdot$GeV data was determined by multiplicity 
cuts (see ref.~\cite{lisa}). 

It is commonly accepted that the kinetic energies of fragments can be 
represented as a sum of thermal and collective components. 
The collective motion (radial flow) is an important ingredient of Au on Au
collisions at intermediate energies and influences strongly the
fragment energies.
Kinetic energy distributions for central events have been analysed in the 
framework of the 'blast model' which is described in detail in 
ref.~\cite{reis1}.
Here, we recall briefly only the main aspects. The collective energy
stored into radial flow was determined by velocity profiles
and an ansatz for the velocity distribution \cite{siem} of the fragments.
The corresponding kinetic energy distributions were reproduced if the 
collective energy
$\varepsilon_{coll}$ amounts to 62$\pm$8 \% of the center-of-mass
energy E$_{C.M.}$ available in the collision. Then, the energy conservation
requires that
\begin{equation}
\varepsilon_{coll} + \varepsilon_{th} = E_{C.M.} + Q 
\end{equation}
where $Q$ is the Q-value of the reaction and $\varepsilon_{th}$ is the
thermal energy. 
In the limit of classical statistics the temperature is
determined by the multiplicities {\em N} of the emitted particles using the
non-relativistic expression
\begin{equation}
\varepsilon_{th} =\frac{3}{2} \cdot (N-1)\cdot T^{\ast}. 
\end{equation}
The temperatures found by this approach are $T^{\ast}$=\\
17.2$\pm$3.4, 26.2$\pm$5.1 and 36.7$\pm$7.5 MeV
for the beam energies 150, 250 and 400 A$\cdot$MeV,
respectively. Below we label them as 'kinetic' temperatures $T_{kin}$. 
It was emphasized that the temperatures $T^{\ast}$ are 'effective'
in the sense that they are not the temperature at freeze-out time, since 
the observed multiplicity may be raised due to late particle decays.
In ref.\cite{reis1}, these
temperatures were used as input for the statistical multifragmentation
models QSM \cite{hahn}, WIX \cite{fai} and SMM \cite{bot1} 
by assuming that the whole entire mass of the colliding Au nuclei 
undergoes thermalization. This assumption fails to reproduce the 
fragment multiplicities. 
The abundances of heavy clusters are underestimated
up to 4 orders of magnitude (e.g. for Z=8 fragments at 400 A$\cdot$MeV.)
In ref.\cite{reis1} it was discussed that
variations of the radial flow energy, the freeze-out densities 
and level density parameters within reasonable physical limits
cannot account for such large deviations. Hence, the question arises
about the applicability of established statistical models to
cluster production in the midrapidity source and the 
meaning of the nuclear temperatures mentioned above.
The present paper is intended
to help to disentangle this 'puzzle'. First we ask whether
these temperatures can be confirmed by other data available 
for the same collision
system Au+Au measured with the same apparatus. Such possibility
offer the yields of hydrogen and helium isotopes \cite{pogg} which
can be treated by the isotope thermometry. 

In ref.~\cite{albe} it was shown that in an equilibrated system
the double-yield ratio ($R_1$/$R_2$) of isotopes is directly related
to the temperature of the corresponding grand canonical ensemble:
\begin{equation} \label{tiso}
T_{iso} = \frac{b}{\ln \left( a \cdot (R_{1}/  R_{2}) \right)}.
\end{equation}
For the consideration of hydrogen ($d,t$) and helium isotopes ($^3He,^4He$)
one needs to fix the parameters $b$=14.32 MeV and $a$=1.59 \cite{moeh}
which include the binding energies,masses and spin degeneracy. 
This isotope thermometer has proved to be successful in many 
applications. In particular, for the first time it was possible to establish 
experimentally the nuclear caloric curve \cite{pochodzalla95}. 

Here, we refer to the isotopic yield ratios and kinetic energy distributions
of $d$, \,$t$,\,$^3He$ and $^4He$ measured with $\Delta E$/$E$ telescopes within the
C.M. polar angle range of $60^0 \leq \Theta_{C.M.} \leq 90^0$ \cite{pogg}.
Most of the projectile fragments are expected to be suppressed within this
angular coverage. The corresponding data sets at 100, 150 and 250 A$\cdot$MeV
were obtained from central event samples selected by the criterion ERAT5 \cite{reis1}.
The integration of the C.M.-
kinetic energy spectra delivered the intensity of deuterons, tritons,
$^3$He and $^4$He from which the ratios R$_1$ and R$_2$ and the 
corresponding isotope temperature T$_{iso}$ were determined.
The obtained values $\langle T_{iso} \rangle =6.34 \pm0.50,\,7.8 \pm0.8$
and $11.51 \pm1.58 MeV$ for 100, 150 and 250 A$\cdot$MeV, respectively,
are about three times smaller than the corresponding kinetic temperatures. 
More details of the distribution of T$_{iso}$ in the 
freeze-out volume are required. Here, we determined from the LCP spectra
presented in ref. \cite{pogg} the isotope temperature in dependence on the velocity.

The C.M. kinetic energies E$_{kin}$ were transformed into the particle C.M. velocities
using the relativistic relation
\begin{equation}
\frac{v}{c} = \frac{\sqrt{E_{kin}(E_{kin} +2mc^{2})}}{E_{kin} + mc^{2}}
\end{equation}
where {\em m} is the corresponding LCP mass, $c$ is the light velocity.
The spectra given in ref.~\cite{pogg} have an equidistant energy binning
for all particles. But, the velocity divisions become 
varying for different particle masses after the abscissa transformation.
In order to cover the same range of velocities we analysed
the kinetic energy distributions to E$_{kin}(d)<$ 110 MeV, 
E$_{kin}(t,^{3}He)<$ 150 MeV and  E$_{kin}(^{4}He)<$ 200 MeV for incident 
energies 100 and 150 A$\cdot$MeV.
The corresponding limits 
are 170, 250 and 300 MeV at 250 A$\cdot$MeV. Equal reference velocities v$_i$ 
for the four LCP's were found by an appropriate interpolation of the 
velocity distributions derived within these limits. The maximum velocity 
up to which yields of all LCP's were available was 
around 0.3$\cdot$$c$ at 250 A$\cdot$MeV
and 0.6$\cdot$$c$ at 1\,A$\cdot$GeV, respectively.  
Then, the isotope temperature $T_{iso}$ was recalculated for subsequent velocities 
v$_{i}$ by means of equation (3) with the ratios
\begin{eqnarray}
R_1= Y(d,v_{i})/Y(t,v_{i})~~~\nonumber \\ 
R_2= Y(^3He,v_{i})/Y(^4He,v_{i}).
\end{eqnarray}

The mean temperatures evaluated within the limited velocity 
ranges are $\langle T_{iso} \rangle =$ 5.30, 6.10 and 9.14$\pm0.60$ MeV, 
respectively, for 100, 150 and 250 A$\cdot$MeV. 
They are smaller than the values derived from the yields 
integrated over the complete energy spectrum, since the contributions 
of fragments with the highest energies are missing. However, this 
treatment is self-consistent for this subset of events, where the collective 
flow dominates over Coulomb and thermal energies and determines mostly the 
fragment velocities. 

As shown in Fig.~1 the temperatures $T_{iso}$ become larger with increasing 
particle velocity, however, they are below the kinetic 
temperatures $T^{*}$ of ref. \cite{reis1}. 
This finding suggests that the obtained data cannot be described by {\em one} 
temperature characterizing the complete equilibrium.
An explanation of this phenomenon could be that fast (preequilibrium) 
nucleons are emitted earlier at very high temperatures and carry away 
excess energy, while slow nucleons form an equilibrium-like source 
\cite{bon2}. 
The observed light fragments are produced during the cooling 
process and the distribution of the chemical temperature versus the fragment 
velocity reflects evolution toward equilibrium. 
Figure~1 shows that in all cases the temperature $T_{iso}$ has some 
saturation at low velocities. 
This could be an evidence for reaching thermalization of
nuclear matter at these velocities. 
Obviously, this thermal
source can be characterized by very moderate {\em chemical} temperatures. 

It is a general observation 
(e.g., see \cite{hxi}) that in intermediate energy collisions 
the kinetic energy spectra of fragments cannot be reproduced by using an unique 
temperature. In the following we propose a model to describe these feature.\\
\begin{figure}
\begin{center}    
\epsfig{file=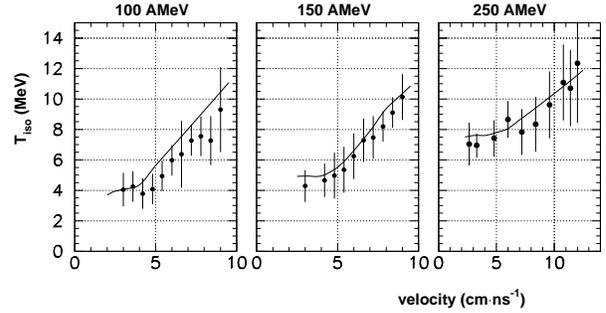,width=9.5cm}
\end{center}
\caption{Isotopic temperatures evaluated from the data in ref.~\cite{pogg} as a function
         of the C.M. velocities of $d$,$t$,$^3He$ and $^4He$. Dots: data. Lines:
         calculations with the code SMMFC including contributions from the
         fast stage (coalescence).    
         }
\label{fig:cool}
\end{figure}

\section{Description of the model}

An appropriate way to describe processes involving 
many particles is the statistical approach. The system characterized
in the initial stage by nonequilibrium distribution 
functions evolves towards equilibration as a result of many interactions
between the
particles. In this process the system runs through different states.
The first one can be considered as 
equilibration of the one-particle degrees of freedom. 
The following evolution toward total thermalization can be 
considered as involving of higher order particle correlations. 
For finite expanding systems the degree of equilibration depends on the 
reaction type. 
It is expected that the equilibration is less effective for nucleons at the surface
than for nucleons deep inside the freeze-out volume. As a result, the mechanisms 
of the fragment production may be different also. 

In the following we are going to interpret the experimental data within the 
framework of the expanded version of the statistical multifragmentation model
\cite{Bon95,bot1} which treats also LCP's emitted prior to the equilibration. 
The model phenomenologically includes collective motion (radial 
flow ($F$)) and, optionally, composite particle production by a new 
coalescence ($C$) algorithm. 
\footnote{henceforth the sign SMMFC denotes this code version} 
The subsequent application of these mechanisms is
aimed at simulating the most important physical processes 
in many-particle systems.

The model parameters have to be derived from fits to the experimental data as
described in section 4. It is important to use 
as many as possible observables for the analysis, and to ensure 
that the deduced parameters and their dependence on the beam energy are 
qualitatively consistent with general predictions of dynamical calculations 
\cite{bon2}. 

Total energy and momentum balance is used for the implemented processes 
(except $\gamma$-emission). First we consider the quantities being constrained. 
A projectile with mass number $A_1$ 
and charge $Z_1$ collides at beam energy $E_{beam}$ with a target ($A_2$, $Z_2$) 
resulting in the center-of-mass energy $E_{C.M.}$ available for the total 
system with \\
$A_0$=$A_1$+$A_2$ and $Z_0$=$Z_1$+$Z_2$. 
Since the equilibrated stage is rather well 
understood, we start the analysis of experimental data related to 
IMF's which are assumed to be produced only in the equilibrated (i.e. thermal) source. 
This source is parametrized by (i) the mass number $A_{s}=A_{rel}\cdot A_0$ 
(where $A_{rel}$ is the relative source size) and corresponding
charge $Z_{s}=A_{s}\cdot (Z_0/A_0)$, (ii) the thermal 
excitation energy $E^{*}$ and (iii) the collective energy per nucleon $E_{flow}$.
In the analysis of the 1 A$\cdot$GeV data the energy released due to pion production
was taken into account. $P_{\pi}$ is the part of $E_{C.M.}$ taken away by pions. 
The remaining matter ($A_{pre}=A_0-A_{s}$, $Z_{pre}=Z_0-Z_{s}$) 
is assumed to be carried away by fast nucleons and  LCP preequilibrium emission. 
In this case the change of the binding energy is given by the
corresponding values for projectile, target, thermal and preequilibrium 
sources: 
$\Delta B = B_1 + B_2 - B_{s} - B_{pre}$.  
From the conservation of the total energy follows 
the energy available for preequilibrium  
emission : $E_{pre} = E_{C.M.} - P_{\pi}\cdot E_{C.M.} - E^{*} - E_{flow} + \Delta B$.\\

\subsection{LCP emission at the fast stage.}

Since the cooling during the expansion process is very fast, 
nucleons have no time to feel the part of the phase space corresponding
to the composite particle production. 
A distribution of nucleons in the phase space at some 'freeze-out' time
is considered as start configuration. 
Generally, any distribution of nucleons in momentum and coordinate space 
after an initial dynamical process is conceivable.
But, in some experiments, e.g. central nucleus--nucleus 
collisions \cite{reis1}, it is possible to select samples which are
nearly isotropic in space and look like thermal events. 
Therefore, in such cases, we can simply assume that the nucleons populate the
available {\em many-body} phase space uniformly, i.e. there is equilibration in 
one-particle degrees of freedom without manifestation of collective  phenomena. 
That gives rise to a thermal distribution 
for individual nucleons in the thermodynamical limit. 
In the calculations we consider the system characterized by $A_{pre}$,
$Z_{pre}$ and $E_{pre}$ and disintegrate the system 
into nucleons by taking away about 7~A$\cdot$MeV (binding energy). 
The remaining 
energy turns into the kinetic energies of nucleons which populate the whole 
available many-body momentum space uniformly. The procedure developed 
in \cite{kopylov} is used to generate the nucleon momenta.

A composite particle can be formed from two or more
nucleons if they are close to each other in 
the phase space. This simple prescription is known as coalescence model.
Here we use the 
coalescence in momentum space which was recently described and
applied in ref.~\cite{neu1}. The basic assumption is that a dynamical 
process, which leads to a momentum redistribution, is very fast (nearly 
instantaneous), so that the coordinates of nucleons are just defined by 
their momenta. This is also justified taking into account quantum properties 
of the system since the wave functions of nucleons can be described in 
momentum space only. 
This type of coalescence model has proven successfully by reproducing 
experimental data (see e.g. \cite{BU63,gutbrod,CS86}).

In the standard formulation of the model it is assumed that the fragment 
density in 
momentum space is proportional to the momentum density of nucleons 
times the probability of 
finding nucleons within a small sphere of the coalescence radius $p_0$. 
For example, in a nonrelativistic approximation, from this hypothesis  
an analytical expression can be derived 
for momentum spectra of coalescent clusters: 
\begin{equation} \label{Ncoal}
\frac{d^3\langle N_{A}\rangle}{d\bar{p}^{3}_{n}} \simeq 
\left( \frac{4\pi}{3} p_{0}^3 \right)^{A-1} 
\left(\frac{d^3 \langle N_{1}\rangle}{d^3 \bar{p}_{n}} \right)^{A} 
\end{equation} 
where $\bar{p}_n $ are the momenta 
per nucleon. $\langle N_{A}\rangle$ and $\langle N_{1}\rangle$ are
the mean multiplicities of 
fragments with the mass numbers $A$ and $1$, respectively.
This equation disregards correlations between different clusters since 
the conservation of the nucleon number is not taken into account. 
Therefore, the above formulae is valid only 
for $\langle N_{1}\rangle \gg$ $\langle N_{2}\rangle \gg$ 
$\langle N_{3}\rangle$ ... 

We developed another formulation of the coalescence model.
Nucleons can produce a cluster 
with mass number $A$ if their momenta relative 
to the center-of-mass moment of the cluster is less than $p_0$. 
Accordingly we take
$|\vec{p}_{i}-\vec{p}_{C.M.}|<p_{0}$ for all $i=1,...,A$, where 
$\vec{p}_{C.M.}=\frac{1}{A}\sum_{i=1}^{A}\vec{p}_{i}$. 
In the following examples the value $p_0\approx94$\,MeV/c 
has been adopted corresponding to relative
velocities $v_{rel}$=0.1$c$ 
in agreement with previous analyses \cite{neu1}. 

In this context we would like to draw attention to a problem
which is sometimes disregarded in these simulations. 
Some nucleons may have such momenta that they can belong to 
different coalescent clusters according to the coalescence criterion. 
In these cases the final decision depends on the sequence of nucleons within
the algorithm. To avoid this uncertainty we developed 
an iterative coalescence procedure. ${\cal M}$ steps are calculated in the coalescence 
routine with the radius $p_{0j}$ which is increased at each step $j$: 
$p_{0j}=(j/{\cal M})\cdot p_0$ ($j=1,...,{\cal M}$). Clusters produced at earlier steps 
participate as a whole in the following steps. In this case the final clusters 
not only meet the coalescence criterion but also the nucleons have the minimum 
distance in the momentum space. This procedure is unique in the limit
${\cal M}\rightarrow\infty$ and we found that in practical calculations it 
is sufficient to use ${\cal M}$=\,5.

The importance of this coalescence mechanism is demonstrated by Fig.~2 which
shows a comparison of measured isotopic yield ratios $d/t$ and $^3He$/$^4He$
in the beam energy range from 35 A$\cdot$MeV to 1 A$\cdot$GeV~\cite{pogg,eos,Xi1,dos1}.
Especially at 
higher incident energies the SMM calculations without consideration of
coalescence underestimate strongly the existing data. The so-called 
'$^3He-^4He$ puzzle' can also be solved by taking into account
the coalescence \cite{neu1}. 

In the calculations we assume that only coalescent particles with $A\leq 4$
are produced. In principle, one can extend the model by considering IMF 
also, though the probabilities of coalescent IMF is small \cite{CS86,neu1}. 
In this case the expected portion of the thermal source will be even 
smaller than with restriction to $A\leq 4$. However, this paper 
is aimed at finding an upper limit for the contribution of thermal IMF's. 
As justification of our assumption we present a good agreement with 
experimental data in section 5. There are also other experimental 
features supporting this assumption. 
In particular, the maximum of IMF production was found at 
small fragment velocities. These velocities correspond to the nearly constant 
temperatures $T_{iso}$~ (see Fig.~1) associated with a thermal-like 
source. 
\begin{figure}
\begin{center}    
\epsfig{file=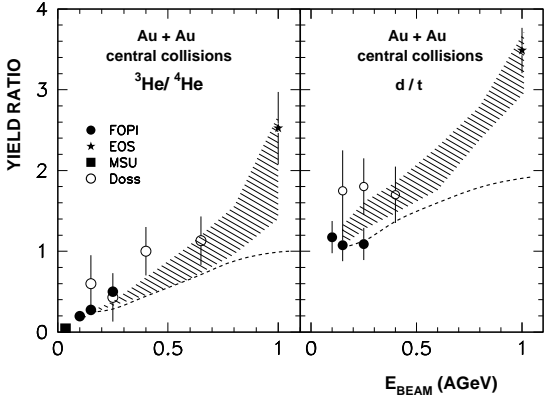,width=8.6cm}
\end{center}
\caption{Isotopic yield ratios of LCP in central Au on Au collisions as a
         function of the beam energy. Data are displayed by symbols:
         FOPI~ \cite{pogg}, EOS~ \cite{eos}, MSU~ \cite{Xi1}, Doss~ \cite{dos1}.
         The hatched areas show SMMFC calculations assuming the uncertainty
         of the input parameters given in table\,1.  
         Dashed lines: Calculations using only the thermal source (without coalescence)
         and the same input (without consideration of uncertainty limits).
         }        
\label{fig:coal}
\end{figure}

\subsection{Thermal source}

At the freeze-out time of several tens of fm/c there is still a lot of 
relative dense nuclear matter in the center of the system in which 
instabilities (usually associated with the liquid-gas type transition) 
may occur. At this stage it is assumed that {\em thermal equilibrium} is 
reached, which is responsible for IMF production.  
The success of the SMM and other statistical models \cite{Bon95,fai,gross} 
to describe such reactions (see e.g. 
refs.\cite{Bon95,agos,Lav,Bouriq,will,neu2}) supports this assumption. 

A microcanonical ensemble of all 
break-up partitions which consist of nucleons and excited fragments of 
different masses \cite{Bon95} is considered.
If $E^{\ast}$ and the volume of the thermal source are fixed,
the statistical weight $W$ of a given break-up partition $j$ (i.e. the number 
of microscopic states leading to this partition) 
is determined by its entropy $S_j$: 
\begin{equation}
W_j^{mic} \propto exp(S_j(E^{\ast},A_s,Z_s)).
\end{equation}
The fragments with 
mass number $A>4$ are treated as heated nuclear liquid drops but the
light fragments with $A\leq 4$ are considered as massive particles
('nuclear gas') having only translational degrees of freedom. The
ensemble of partitions is generated by Monte--Carlo methods according to
their statistical weights $W_j$ constrained by the conservation laws.
The microcanonical temperature $T_{th}$ is found from the energy balance by 
taking into account 
the Coulomb interaction, binding energies and excitations of fragments. 
After break-up of the system
the fragments propagate independently in their mutual Coulomb fields and
undergo secondary decays. The deexcitation of large fragments
($A>16$) is described by the evaporation-fission model, and
for smaller fragments by the Fermi break-up model \cite{bot1,fermi}.

\subsection{Relation between coalescence mechanism and thermal fragment 
production.}

An important relationship can be formally established \\
between the
coalescence and thermal models as far as fragment production is considered.
This will be illustrated by a simple statistical case.
A system with the total number of 
nucleons $A_0$ disintegrates into fragments with mass number $A$, which are 
characterized only by their binding energy $B_A$. The fragments are 
considered as Boltzmann particles moving without interaction in the 
volume $V$. Then the statistical partition sum can be calculated as 
\begin{equation} 
\eta=\sum_{partitions} \left(
\prod_{A}\frac{V}{(2\pi \hbar)^3} \int d^3 \bar{p}
\cdot e^{-\frac{\bar{p}^2}{2 m_{n} A T}}\cdot P_A \right), 
\end{equation} 
where $T$ is the kinetic temperature characterizing the fragment 
translational motion, 
$m_{n} \approx 0.94$\,GeV is the nucleon mass 
and $\bar{p}$ is the fragment momentum. The product 
includes all fragments in a partition. The momentum and the center of 
mass conservations are disregarded. 
The magnitude $P_A$ is proportional to the probability 
of the formation of fragment $A$, which can be written as $P_A=exp(-B_A/T)$
in the case of full equilibration characterized by the 
canonical temperature  $T$. However, the 
process of fragment formation may be complicated, and this 
process may not be related only to the thermal kinetic motion. 
Generally, one can write 
$P_A=exp(-B_A/T_A)$, where $T_A$ is a parameter ('temperature') related to 
a given fragment. 
One can introduce a Lagrange multiplier $\mu$ (like a 'chemical potential'), which 
can be found from the condition $\sum_{A}\langle N_{A}\rangle A=A_{0}$ (see ref.~
\cite{Bon95}). Then the partition sum can be calculated as 
\begin{equation} 
\eta=\sum_{N_{1}=0}^{\infty}\cdots\sum_{N_{A_{0}}=0}^{\infty}
\prod_{A}\left[\frac{\langle N_{A}\rangle^{N_{A}}}{N_{A}!}\right].
\end{equation} 
Here $N_{A}$ is the multiplicity of fragment $A$ in a partition, and  
\begin{equation} 
\langle N_{A}\rangle=
\frac{V}{\lambda_{T}^{3}}A^{3/2}\exp\left[-\frac{B_A}{T_A}+\mu A\right], 
\end{equation} 
with the thermal wavelength 
$\lambda_{T}=\left(2\pi\hbar^{2}/m_{n}T\right)^{1/2}$. The last equation 
can be rewritten as 
\begin{equation} \label{Nt}
\langle N_{A}\rangle=\langle N_{1}\rangle^A
\left(\frac{\lambda_{T}^{3}}{V}\right)^{A-1}A^{3/2}\exp(-\frac{B_A}{T_A}). 
\end{equation} 

At this point we can establish a correspondence \mbox{between} the thermal and 
the coalescence models (see also \cite{mek}). 
In the coalescence model, the fragment multiplicity can be determined 
after integration of equation (\ref{Ncoal}). If the nucleons are assumed to 
have a Maxwell--Boltzmann distribution of the same kinetic 
temperature one can easily get 
\begin{equation} \label{Nc}
\langle N_{A}\rangle \simeq \left(\frac{4\pi}{3}p_0^3\right)^{A-1}
\frac{\langle N_1\rangle ^A}{(2\pi m_{n}T)^{3/2(A-1)}A^{3/2}}.
\end{equation} 
By comparing equations (\ref{Nc}) and (\ref{Nt}) we obtain 
a formal relation between the model parameters:
\begin{equation} \label{connec}
\frac{4\pi p_0^{3}V}{3h^{3}} \simeq 
\left( A^{3} \cdot exp(-\frac{B_A}{T_A})\right)^{1/(A-1)} .
\end{equation} 
A possible physical interpretation of this relation is the following. 
If we assume that the freeze-out density and the coalescence parameter are 
determined by a short range interaction between nucleons and properties 
of formed fragments, then there is an effective 'temperature' $T_A$ which 
characterizes the produced coalescent fragments. In the case of 
saturation of the binding energy, 
i.e. $B_A\sim A$, the effective temperatures 
do not differ much for fragments with different $A$. 
That resembles the chemical temperature, when the relative 
probabilities of different fragments are determined by this temperature. 
One can see that the isotope temperature defined by equation (\ref{tiso}) 
corresponds exactly to this temperature. 
Moreover, one can also use relation (\ref{connec}) in another way,
namely, one can determine the coalescence parameters $p_0$ 
for different fragments from the 
experimentally obtained chemical temperatures \cite{cibor}.
Therefore, if statistical models are applied to interpret 
processes with strong dynamical features, 
we should take into account that the fragment velocities may be 
determined by the kinetic temperature (or by the initial dynamics) 
but not by the chemical temperature.

\subsection{Radial flow}

After a central collision the collective expansion is assumed to change 
only the velocity of the fragments taken into account by a 
flow velocity profile $\vec{v}_f(r)=(\vec{r}/ R) \cdot v_0$ 
proportional to the position $\vec{r}$ from the center of the disassembling 
equilibrated source \cite{Bon95}. 
Within this scenario the flow velocity is superimposed onto 
the stochastic motion of the generated fragments 
and only the stochastic thermal part is 
responsible for fragment production. The radial flow is supposed to 
change the fragment velocities but not the fragment yields. 
This ansatz comes from the hydrodynamical picture, however, in the case of 
nuclear multifragmentation there are theoretical arguments for using 
this approach at 
$E_{flow} \leq 3$ A$\cdot$MeV \cite{Bon95,bon4}. In this paper we extend 
it also to higher flow energies in order to find 
limits of this approach by analyzing the experimental data. 
The adequacy of the above ansatz 
is formally supported by statistical-like properties 
of fragments produced in a dynamical process such as coalescence. 
The lattice model calculations \cite{lattice} 
show also that the flow may not influence statistical 
fragment formation. 

This hypothesis of decoupling thermal and collective motions 
is sufficient for a reasonable reproduction of experimental data 
\cite{Lav,Bouriq,will,neu2}. 
By introducing the phenomenological radial flow profile we fit the IMF 
velocities, however, we do not explain the 
velocities. Nevertheless, since we conserve the total energy 
and momentum in the system, this receipt allows to simulate 
individual multifragmentation events and to compare them directly 
with experimental data.\\

The described modified statistical model does not pretend to be a complete 
substitute for dynamical calculations 
but it should be considered as a effective tool for 
a primary analysis of experimental data. Then, the observables being properly 
reproduced, the physical meaning of the fitted parameters can be interpreted. 
The knowledge of these parameters (in particular, the temperatures) 
is supposed to be important for applications of thermal descriptions 
for many nuclear processes. 
The code SMMFC allows us to perform calculations with high statistics for
large systems, as Au on Au, with reasonable computing time. The code  
produces event distributions directly related to the observables, e.g. the 
multiplicity of an event as well as the charge, the mass, the kinetic 
energy and the polar and azimuthal angles. In this way, 
the generation of single events allow us to process the calculated
quantities in a way like the experimental data samples. Our implemented 
filter permits to study the influence of the detector geometry and resolution, the
Z-dependent registration thresholds and chosen cut conditions.\\

\section{Adjustment of the model parameters}

The following four parameters of the model (see \mbox{section 3}) 
\begin{center} 
$E^{*}$,\,$A_{rel}$,\,$E_{flow}$ and $P_{\pi}$
\end{center}
determine the fragment production. The influence of the break-up density
$\varrho$ is discussed below.
A scanning over the complete parameter space, used in \cite{bon2,Des}, 
is too time-consuming in our case so that 
we disentangle this coupled parameter set by
finding specific sensitivities of the parameters to certain observables.
We refer in the following only to the key observables which are
necessary to fix the model parameters.\\  
\begin{figure}
\begin{center}    
\epsfig{file=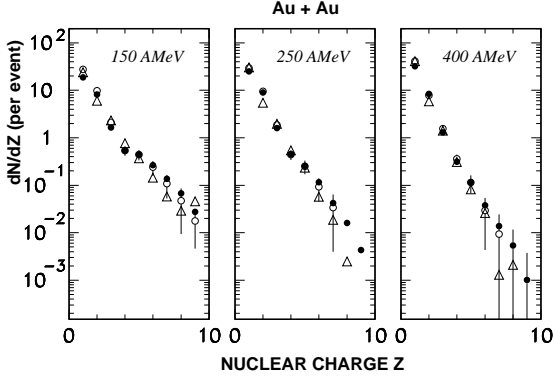,width=8.6cm}
\end{center}
\caption{Measured and simulated multiplicities N of nuclear charges. 
         Black dots: data from ref.~\cite{reis1}.
         Open triangles: subset of data analysed by the selection criterion
         $ERAT \geq 0.7$ and $\Theta_{C.M.} \geq 25^{0}$ within the plastic wall acceptance,
         Open circles: filtered model calculations with input parameters from table 1. 
         The calculated Z=1,\,2 multiplicities contain contents from both the 
         preequilibrium and thermalized sources.}
\label{fig:comp}
\end{figure}

\subsection{Thermal excitation energy}

Multiplicity distributions of charged particles from central Au+Au collisions 
(ref. \cite{reis1}) are plotted in Fig.~3.
In addition, our results obtained from a subset of data
selected by the centrality criterion $ERAT\geq\ 0.7$ and $\Theta_{C.M.}\geq 25^{0}$
are also shown in this figure.
Since the experimental distributions have a nearly exponential shape 
dN/dZ$\propto exp(-\alpha Z)$
over the whole range of measured charges, they can be fit by an exponential form.
We included into the fits only 
the part of the charge distributions with $Z \geq 3$ to exclude the influence of
LCP on the parameter $\alpha$. In ref.~\cite{reis1} also the Z=4 points were excluded
from the fits performed within $3 \leq Z \leq 10$. Our fit parameters 
$\alpha (3 \leq Z \leq 6)$
obtained from the mentioned subset are somewhat larger than the corresponding
one's of ref.~\cite{reis1}. Since the errors $\Delta \alpha$ from each fit are
small compared to possible disturbance of the charge distribution  by evaporative 
processes and sequential decays, we used the difference between the fit values 
given in ref.~\cite{reis1} and $\alpha (3 \leq Z \leq 6)$ as a measure of 
possible deviations. 

The parameters $\alpha$ derived from the data are nearly proportional to the 
available center-of-mass energy. Otherwords, 
the calculated charge distributions show that the steepness parameter 
$\alpha (3 \leq Z \leq 6)$  increases also linearly with the 
excitation energy of the thermal source. In order to find a quantitative 
relation, numerous charge distributions were simulated at various 
excitation energies. From a linear fit within the limits 3$\leq$Z$\leq$6  
we found the relation $\alpha$ = $-0.571 + 0.1508 \cdot E^{\ast}$ 
within the considered ranges of the source parameters. 
This result is nearly independent on the source size A$_{rel}$, freeze-out
density $\varrho$ and the radial flow E$_{flow}$.
Steepness parameters $\alpha$ obtained from calculated charge yields for the
full phase space deviate only by a
few percents from those which were processed by the filter routine
implemented in SMMFC. 

The procedure of determining the parameters at 250 A$\cdot$MeV is 
illustrated in Fig.~4. 
The values $\alpha$=0.91 and $\alpha$=1.15 \cite{reis1} 
found from fits to the data determine the interval of the corresponding 
excitation energy E$^{\ast}$~ (upper panel of Fig.~4).

\subsection{Size of the thermal source}

The interval of excitation energies being fixed, we are going to estimate
the relative source size $A_{rel}$ by means of the IMF multiplicity 
which depends in terms of SMMFC on both $A_{rel}$ and $E^{\ast}$.
The lower panel of Fig.~4. demonstrates how 
$A_{rel}$ is estimated from the overlap of the experimental IMF multiplicity 
\cite{reis1} 
with the calculated one's. $A_{rel}= 0.50 \pm 0.05$ at 250 A$\cdot$MeV 
is consistent with both the multiplicity data and the estimated limits of 
$E^{\ast}$. The expected sizes of the equilibrated source at 150 A$\cdot$MeV 
and 400 A$\cdot$MeV 
have been obtained analogously and the results are given in table 1.
These parameters are consistent with those extracted in ref.~\cite{Lav}
at somewhat lower incident energies. The obtained trend of the
decreasing size of the thermal source with the beam energy is supported by 
dynamical calculations. For example, the analysis of the FOPI data with 
the hybrid model {\em BUU+SMM} of ref.~\cite{heid} has shown that in central 
Au+Au collisions the 
fraction of thermalized matter drops from $A_{rel}$=0.48 at 150 A$\cdot$MeV 
to $A_{rel}$=0.3 at 250 A$\cdot$MeV, respectively.\\

\begin{figure}
\begin{center}    
\epsfig{file=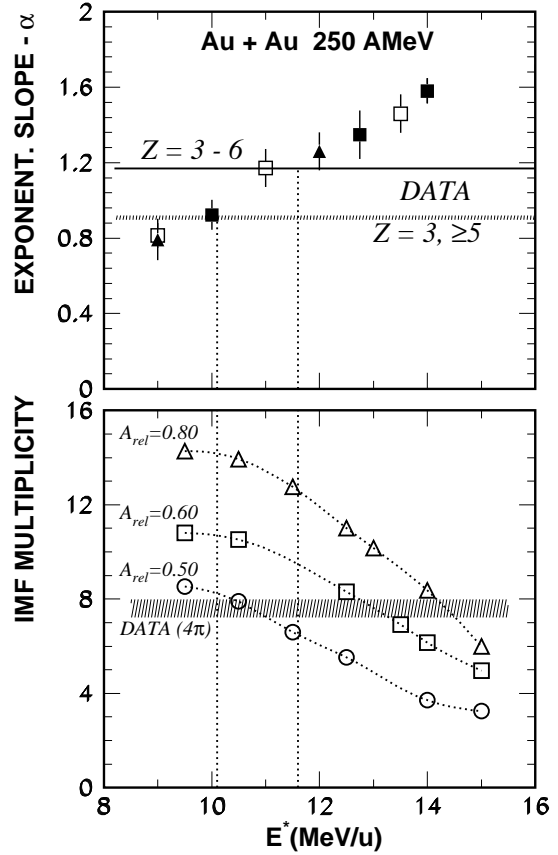,width=8.6cm}
\end{center}
\caption[   ]{Determination of the input parameters of the model.
         Upper part: steepness parameter $\alpha$ vs. excitation energy 
         E$^{\star}$.
         Dashed horizontal lines: data taken from \cite{reis1} and 
         this work.
         SMMFC calculations are denoted by symbols. Full squares:
         E$_{flow}$ =0.55 $\cdot$ E$_{C.M.}$, A$_{rel}$=0.3.
         Black triangles: E$_{flow}$ =0.55 $\cdot$ E$_{C.M.}$, A$_{rel}$ =0.5,
         Open squares: E$_{flow}$ =0.37 $\cdot$ E$_{C.M.}$, A$_{rel}$=0.3.
         Lower part: IMF multiplicity (3$\leq Z \leq$ 6) vs. excitation energy.
         Hatched area: data from ref.\cite{reis1}. The symbols are SMMFC
         calculations for three sets of A$_{rel}$ at fixed parameters
         E$^{\star}$ =11.0 MeV and E$_{flow}$ =0.55 $\cdot$ E$_{C.M.}$. The 
         estimated range of E$^{\star}$ is between the dotted vertical lines.}         
\label{fig:parameter}
\end{figure}
\begin{table}
\caption{Properties of the thermal sources. Deviations of the parameters
given in parentheses are admissible to reproduce the experimental data
within the error bars.E$_{flow}$ is used in the model calculations.
The parameters for 100 A$\cdot$MeV and 1 A$\cdot$GeV are extrapolated one's.}
\label{tab:par}
\begin{center}
\begin{tabular}{|l c c c c |}  
\hline
$E_{beam}$    &$E^\ast$      &  $T_{th}$ &   $A_{rel}$    & $E_{flow}$ \\

(A$\cdot$MeV )&(A$\cdot$MeV) &(MeV)      &                &(A$\cdot$MeV)\\
\hline
100          &$\simeq$9.0   & $\simeq$7.3 &   $\simeq$0.81 &  10.5(1.5)\\
\hline
150          &10.0(1.1)      &7.8(0.4)   &  0.72(0.11)    &  20.(3.0) \\
\hline
250          &10.8(0.8)      &8.4(1.1)   &  0.50(0.05)    & 32.(6.)\\
\hline
400          &15.0(1.4)      &11.5(2.4)  &  0.50(0.13)    & 56.8(6.)\\   
\hline
1050         &24.0(2.0)      &21.2(5.0)  &  0.26(0.10)    & 81.(10.) \\
\hline
\end{tabular}
\end{center}
\end{table}
\begin{table}
\caption{Reproduction of IMF multiplicities.}
\begin{center}
\begin{tabular}{|c  c  c|}
\hline
$E_{beam}$ &$\langle IMF \rangle$ &  $\langle IMF \rangle$ \\
(A$\cdot$MeV)    &      SMMFC$(4\pi)$     & data$(4\pi)$     \\
\hline
150        &   11.61              &10.35$\pm$0.06\,\cite{reis1},\,12.6\,\cite{kuhn}\\
\hline
250        &   8.42               &7.58$\pm$0.04\,\cite{reis1},\,8.3\,\cite{kuhn}\\
\hline
400        &   5.50               & 5.16$\pm$0.03\,\cite{reis1},\,5.7\,\cite{kuhn}\\
\hline
\end{tabular}
\end{center}
\end{table}

\subsection{Radial flow}

As start parameter we used results from previous analyses 
\cite{reis1,pogg,lisa,petro} 
which were slightly varied to get an optimum reproduction of 
the fragment's mean kinetic energies. The values $E_{flow}$ and the allowed 
spread given in table~1 provide a satisfactory agreement between data and 
calculation. These values are in agreement with other analyses of the 
data \cite{reis3}.

\subsection{Energy release by pions}

The quantity $P_{\pi}$ was estimated (i) from pion multiplicities 
predicted by Vlassov-Uehling-Uhlenbeck transport model calculations in dependence of
the beam energy at impact parameters of b=3 fm \cite{stoe}, 
(ii) from the $\Delta$-resonance production probability in central Au+Au collisions \cite{aver},
(iii) from the pion-to-proton ratios at 1.05 A$\cdot$GeV \cite{pelt} and (iiii)
the data in ref.\,\cite{munz}. From all those references one can conclude
that pion production
at 1.05 A$\cdot$GeV carries away  $\simeq $10 \% of the available C.M. 
energy. For the lower considered energies the pion contribution is  
neglegible.

\subsection{Break-up density}

In the model SMM/SMMFC the free volume influencing the translational 
entropy of partitions is not determined by the total volume (or density) of 
the system, though this assumption is adopted in some other statistical 
analyses \cite{kuhn}. 
The free volume reflects dynamics of fragment formation and it depends on 
the fragment multiplicity \cite{Bon95}. This ansatz is important for a 
good reproduction of experimental data \cite{Schar}. In the model the density 
influences directly only the Coulomb interaction in the system. 
The resulting excitation energies of the equilibrated source 
are large compared to the Coulomb energies so that
only minor changes in the fragment yields are expected if the freeze-out density is changed.
This was confirmed by corresponding calculations 
within $1/6 \leq \varrho$/$\varrho_0 \leq 1/3$, where $\varrho_0$ is the 
normal nuclear density.  
Our final calculations were performed with a freeze-out density of
$\varrho$=1/6 $\cdot \varrho_0$.

\section{Comparison of experimental data with model calculations and discussion}

\subsection{Fragment charge distributions and multiplicities}

Since our input parameters for SMMFC have been derived only from the IMF observables
it is important to proof to what extent the model reproduces also the LCP multiplicities.
Figure~3 and table 2 show that the
model is able to reproduce the charge distributions from LCP's to IMF's in shape as well as
in absolute scale provided that preequilibrium nucleons and the coalescenced particles
are taken into account. On the contrary, calculations performed without coalescence
show a clear underestimation of the Z=2 multiplicity as seen by comparison of 
tables 3\,and\,4 based on the data taken from Ref.~\cite{reis1}. 
\begin{table}
\caption{Calculated multiplicities of LCP's in central Au+Au collisions using the
scenario (ii) and the parameters from table 1. 
Typical uncertainties in the resulting sum are $\pm$ 5.}

\label{tab:mult}
\begin{center}
\begin{tabular}{|c|c|c|c|c|c|}
\hline
 &E(A$\cdot$MeV)&pre&equ&sum&data~\cite{reis1}\\
\hline
{\em Z}=1 &150     &32.2  &32.4  & 64.6       &61.84(0.58)\\ 
\hline
{\em Z}=1 &250     &52.2  &30.8  & 83.0       &75.82(0.62)\\ 
\hline
{\em Z}=1 &400     &60.0  &34.8  & 94.8       &92.04(0.62)\\
\hline\hline
{\em Z}=2 &150     &5.3   &15.6  & 20.9       &26.76(0.36)\\
\hline
{\em Z}=2 &250     &8.9   &14.5  & 23.4       &27.27(0.36)\\
\hline
{\em Z}=2 &400     &9.3   &14.3  & 23.6       &24.16(0.30)\\

\hline
\end{tabular}
\end{center}
\end{table}
\begin{table}
\caption{Same as table 3, but calculations without coalescence.}

\label{tab:mult}
\begin{center}
\begin{tabular}{|c|c|c|c|c|c|}
\hline
 &E(A$\cdot$MeV)&pre&equ&sum&data~\cite{reis1}\\
\hline
{\em Z}=1 &150     &40.6  &32.4  & 73.0  &61.84(0.58)\\ 
\hline
{\em Z}=1 &250     &69.0  &30.8  & 99.8  &75.82(0.62)\\ 
\hline
{\em Z}=1 &400     &77.8  &34.8  & 112.6 &92.04(0.62)\\
\hline\hline
{\em Z}=2 &150     &1.4   &15.6  & 17.0  &26.76(0.36)\\
\hline
{\em Z}=2 &250     &0.6   &14.5  & 15.1  &27.27(0.36)\\
\hline
{\em Z}=2 &400     &.1   &14.3   & 14.4  &24.16(0.30)\\
\hline
\end{tabular}
\end{center}
\end{table}

\subsection{Isotope Temperatures T$_{iso}$}

In order to describe the experimental findings shown in Fig.~1 
we performed model calculations 
by using the parameters evaluated in section 4 (see \mbox{table~1}). The 
calculation at 100 A$\cdot$MeV was performed with extrapolated parameters 
given in table 1, which are consistent with parameters extracted in 
ref.~\cite{Lav}. 
The generated kinetic energy distributions of $d$,\,$t$,\,$^3He$ and 
$^4He$ were filtered by the cuts set in the experiment \cite{pogg}. 
Then the calculated LCP spectra were 
transformed into velocity distributions like the experimental data. 
The temperatures T$_{H-He}$ obtained from that reproduce almost 
quantitatively the data (see Fig.~1.) 

In the calculations the increase of T$_{iso}$ with velocity is 
mainly caused 
by a combined effect of the thermal and coalescence mechanisms, since 
they favour production of LCP's with different energies \cite{neu1}. 
The temperature in this range of kinetic 
energies is also sensitive to the 
radial flow: the simulations undershoot the data if too less radial flow 
is assumed and 
overshoot them for too much flow. The parameters within the limits 
given in \mbox{table~1} allow a reasonable reproduction of the data.

The isotope temperatures T$_{iso}$ obtained from energy-\\integrated 
yields, as well as from the yields in the limited range of 
fragment velocities 
(see section 2) are presented in Fig.~5. This figure shows also 
results of the ALADIN collaboration obtained for central Au+Au collisions 
in the energy range from 50 A$\cdot$MeV to 
200 A$\cdot$MeV~\cite{serf}, as well as data at 35 A$\cdot$MeV~\cite{huan}, 
which match also the found trend. Figure~5 shows also the microcanonical 
temperature  $T_{th}$ of the thermal source calculated with the code SMMFC. 
As discussed in \cite{bon5} this temperature is slightly different 
from T$_{H-He}$, however it shows clearly the same behaviour with 
increasing beam or excitation energy.
Therefore, the isotope temperature at low beam energies can be used 
to deduce the temperature of the thermal source 
\cite{bon5}. 

\begin{figure}
\begin{center}
\epsfig{file=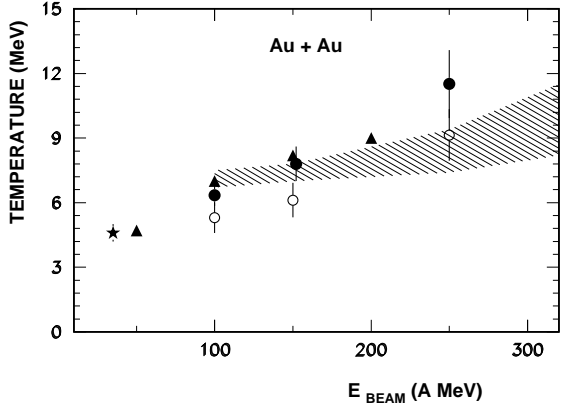,width=8.6cm}
\end{center}
\caption{Temperatures as a function of the beam energy.
         Dots: isotope temperatures obtained from the yield ratios $d/t$ and 
         $^3He/^4He$  
         (integrated over the particles kin. energies)\cite{pogg,eos},
         open dots: the same but in the velocity range limited by 
         $v/c \approx$ 0.3, 
         black triangles: the same, taken from \cite{serf},
         black asterisk: temperature from ref.~\cite{huan},
         hatched area: limits of the microcanonical temperature calculated 
         with SMMFC using the input parameters of table 1.}
\label{fig:temper}
\end{figure} 

\subsection{Kinetic energy spectra}

In this section we compare measured kinetic energy distributions of 
fragments produced at 250 A$\cdot$MeV and \mbox{1~A$\cdot$GeV} with 
corresponding calculations. Calculations with only the thermal source fail 
to reproduce the tails of the spectral shapes of Z=1 and Z=2 particles at 
250 A$\cdot$MeV~(see Fig.~6). A consideration of the preequilibrium 
contribution improves the calculated shape. The displayed two data sets 
of Z=1 distributions demonstrate that the spectral shape is rather 
sensitive to the criterion how central events are selected. The 
calculated SMMFC distribution is close to the event sample obtained by 
the stringent combined criterion of ERAT and directivity 
(see ref.\cite{reis1}).

The disagreement of the tails in the proton spectra might be due to 
deviations of the initial distributions of protons from a 
one--particle equilibration. Neither the spectral shape nor the 
multiplicity of light clusters can be reproduced without consideration 
of coalescence.

Generally, the presence of the radial flow affects the energy
spectra and imitates a high temperature. 
As supposed, the radial velocities of the fragments depend on their 
positions in the freeze-out volume. This leads to an additional 
difference between velocities of fragments 
which is not connected with their thermal random 
motion. Therefore, the deduced source temperature may be essentially lower 
than the temperature extracted from fits of kinetic energy distributions to 
the data. 
For example, the calculations  shown in Fig.~6 were performed with radial 
flow of E$_{flow}$ = 32 A$\cdot$MeV. 
The spectra shapes of lithium clusters generated by a 
thermal distribution 
superimposed with this flow are close to the measured 
data. The Siemens-Rasmussen formula \cite{siem}, which does not 
take into account the effect of the fragment positions, 
reproduces the shape of this distribution also, if 
the kinetic temperature is $T_{kin} \simeq $ 29 MeV. 
However, this temperature is considerably larger than the thermal one. 

In Fig.~7 we compare the mean kinetic energies calculated also for the 
heavier clusters 
with the corresponding data~\cite{petro}. The satisfactory agreement between 
data and calculation is consistent with the result 
of ref.~\cite{reis1} where it was found that 34.0 $\pm$
3.9 A$\cdot$MeV of the available C.M. energy is stored into the radial flow.

\begin{figure}
\begin{center}    
\epsfig{file=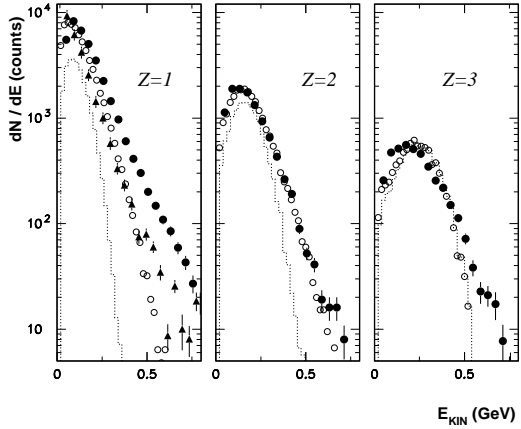,width=9.1cm}
\end{center}
\caption{Kinetic energy spectra of ejectiles with charges Z=1,\,2,\,3 at 
        250 A$\cdot$MeV. Full dots: experimental data for the cut 
        $\Theta_{C.M.} \geq 25^{0}$ and $ERAT \geq 0.7$, full triangles:
        data from ref.\cite{reis1}.
        Dashed histogram: thermal spectrum, open circles:
        sum of thermal and preequilibrium parts (phase space generation,
        coalescence included). Calculations and data are scaled among
        themselves to compare the spectral shapes.}
\label{fig:ekin}
\end{figure}
\begin{figure}
\begin{center}    
\epsfig{file=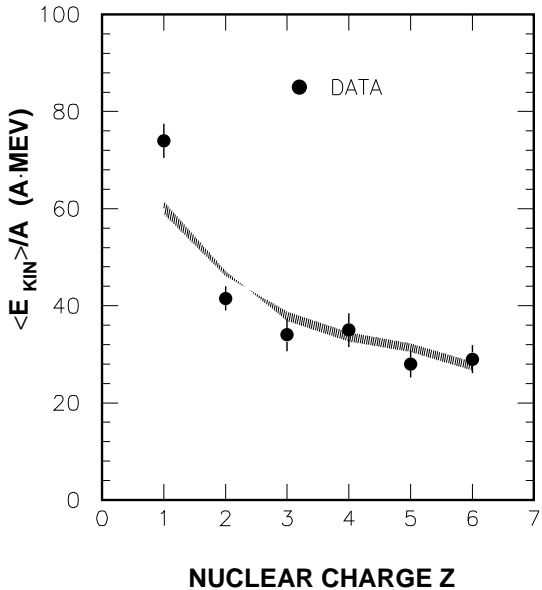,width=8.6cm}
\end{center}
\caption{Mean kinetic energies per nucleon at 250 A$\cdot$MeV. Dots: data from ref.\,\cite{petro}
         for the azimuthal angle $\varphi$=180$^o$. 
         Hatched area : calculations with 32$\leq E_{flow} \leq$ 34 A$\cdot$MeV.}
\label{fig:flow}
\end{figure}
\begin{figure}
\begin{center}
\epsfig{file=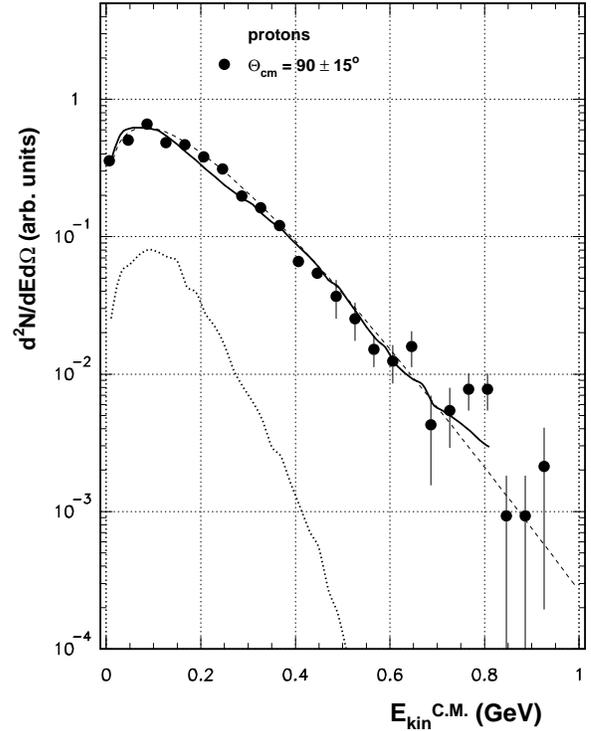,width=8.6cm}
\end{center}
\caption{Differential kinetic energy distributions of protons at 1.0 A$\cdot$GeV. 
Dots: EOS data~\cite{lisa}, dotted line: calculated spectrum 
of the thermal source, solid line:
sum of the fast-stage 
and thermal sources normalized to the data, dashed line: fit of the
calculated proton distribution
using the Siemens-Rasmussen relation. 
}
\label{fig:protons_1gev}
\end{figure} 

Next we applied SMMFC to analyse 
data obtained in central Au+Au collisions around 1~A$\cdot$GeV 
\cite{eos,lisa}. At this incident energy mostly LCP's are produced 
and the influence of the thermal source on the fragment production 
is very limited. 
The model input parameters  E$^*$ and A$_{rel}$ (given in table 1) 
were estimated from the values at low $E_{beam}$ 
by straightforward extrapolation. 
The energy stored in radial flow was taken from ref.~\cite{lisa}. 
A slight variation 
of the parameters ($\pm$12\%) was allowed to find optional agreement 
with the data.  
Proceeding on this input, the model calculations
\footnote
{no filter was applied to the calculated events since it was emphasized (ref.\cite{eos}) that
'the TPC...allows measurements of spectra up to angles of $90^o$ in the center of mass
with $no$ low-$p_T$ cut for central events'.} 
give a quite reasonable description of the proton distribution (Fig.~8),  
demonstrating the high degree of one-particle equilibration 
reached in this central event sample. 
\begin{figure}
\begin{center}
\epsfig{file=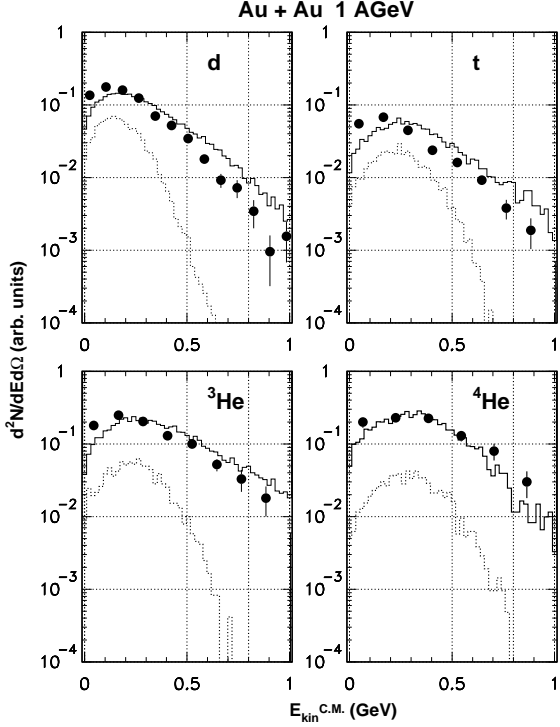,width=8.6cm}
\end{center}
\caption{Energy distributions of hydrogen and helium isotopes. Black dots:
         data taken from ref.~\cite{lisa}. Solid histograms: sum of the fast-stage
         and thermal sources. Input E$_{flow}$ =90 A$\cdot$MeV, A$_{rel}$=0.26,
         P$_{\pi} \simeq $0.10. Dashed histograms: the same input, but without
         coalescence. Calculations are normalized to the data.
         } 
\label{fig:fourspectra}
\end{figure}
However, protons from the complete thermalized source contribute only with 
a very small fraction (see the dotted line in Fig.~8). 
In tables~5 and 6 we compare experimental and calculated multiplicities 
and average kinetic energies of LCP's. A sufficient agreement between 
data and calculations could be achieved. 
The data given in table\,5 were used to calculate the 
temperature T$_{H-He}$ = 18.2$\pm$5.9 MeV. 
The corresponding isotope temperature calculated with SMMFC amounts 
to T$_{iso} \simeq $ 18.0 MeV. However, these 
temperatures are considerably lower than the kinetic temperatures 
determined by the slope of the energy spectra. 
\begin{table}
\caption{Multiplicities of LCP's in central Au+Au collisions at 1.0 A$\cdot$GeV.
    Upper row: data \cite{eos,scott}, lower row: SMMFC calculations.}
\begin{center}
\begin{tabular}{|c|c|c|c|c|}
\hline
 protons        & deuterons      & tritons       & $^3$He        & $^4$He       \\
\hline      
78.82          & 33.34          & 9.56          & 6.06          & 2.40         \\
$\pm$1.48       & $\pm$0.98      & $\pm$0.48     & $\pm$0.4      & $\pm$0.26    \\
\hline
 84.7           & 35.9           & 9.6           & 9.7           & 3.6          \\  
\hline
\end{tabular}
\end{center}
\end{table}
\begin{table}
\caption{Average kinetic energies $<E_{kin}>$ in central Au+Au collisions at 1.15 A$\cdot$GeV.
Upper row: data \cite{lisa}, lower row: SMMFC calculations.}
\begin{center}
\begin{tabular}{|l|c|c|c|c|}
\hline
 protons      & deuterons      & tritons        & $^3$He         & $^4$He      \\
\hline  
203$\pm$3     & 263$\pm$3      & 322$\pm$12     & 328$\pm$12     & 359$\pm$20   \\
  MeV         &  MeV           &   MeV          &     MeV        &   MeV        \\   
\hline 
233$\pm$11 & 278$\pm$18 & 331$\pm$12 & 307$\pm$15 & 317$\pm$15      \\
  MeV      &    MeV     &   MeV      &    MeV     &  MeV            \\
\hline 
\end{tabular}
\end{center}
\end{table}

In order to illustrate this point we compare our results with the fit 
of the data performed in \cite{eos,lisa} based on the blast scenario 
of Siemens and Rasmussen \cite{siem}. 
In Fig.~8 we show also the corresponding blast model fit to the proton 
spectrum equivalent to the high kinetic temperature of 
T$_{kin}$=81 MeV~\cite{lisa}. Nearly the same temperature reproduces 
also the $d,\,t,^3He$ and $^4He$ distributions \cite{lisa}.
The obvious difference between kinetic and isotope temperatures 
suggests that each of both temperature evaluations is related to different 
processes. 

In Fig.~9. we compare 
the differential LCP spectra taken from ref.~\cite{lisa} with 
SMMFC calculations using identical input parameters. 
A reasonable simultaneous reproduction of the spectral shapes of the 
above species can be achieved, if coalescence is taken into account. 
Results without coalescence (dashed histograms) deliver shapes 
quite different from the data. The shapes of LCP energy distributions 
calculated by SMMFC including coalescence are similar to 
the corresponding one's obtained by the blast-model fits. From that 
one can conclude that the kinetic temperature, evaluated from such fits, 
characterizes the initial distribution of nucleons but it 
is different from the isotope temperature characterizing the
chemical composition of the produced fragments.

Using the procedure described in section 2 we obtained 
the isotope temperature $T_{H-He}$ as function of the radial velocities. 
However, contrary to the increasing trend 
shown in Fig.~1, at 1\,A$\cdot$GeV beam energy the temperature $T_{H-He}$ 
does not increase with increasing velocity. Probably, this is a consequence
of the coalescence mechanism which is mainly responsible for the 
LCP production (see also discussion in section 5.4).

\subsection{Correlation of the radial flow with the chemical temperature} 

By comparing Fig.~5 with table 1 one can see that both the isotope 
temperature, characterizing the produced fragments, and the radial flow, 
reflecting the dynamics of the process, increase with the beam energy. We 
suggest that the correlation between these parameters may provide  
complementary information on mechanisms of fragment production. 

As pointed out, there are two contributions to the isotope 
temperature given by the SMMFC. 
The first one is related to the fragment production in the thermal 
source, which dominates at low flow energies. 
Experimentally, this temperature could be identified approximately
at small flow velocities, as seen in Fig.~1. 
In the following we depict $T_{H-He}$ extracted at fragment velocities 
$v\leq$5 cm/ns as the 'thermal' isotope temperature.   
Contrary to this temperature we call isotope temperatures
obtained from energy-integrated yields 'total' ones. 
Figure~10 presents a phenomenological 
relation between the isotope temperature and the radial flow energy 
of the thermal sources. 
The 'thermal' isotope temperatures, shown in Fig.~10 as open circles, 
increase with increasing radial flow. 
With regard to other experimental data we note that the isotope temperature 
corresponding to central Au on Au collisions at 35 A$\cdot$MeV~\cite{huan} 
should be considered also as the 'thermal' one, since the dominating 
thermal source includes the small radial 
flow energy of $\leq $ 1 A$\cdot$MeV\cite{agos}.\\ 

The second contribution to the isotope temperature comes from the 
coalescence mechanism and it provides higher values of the 
temperature. In order to compare both contributions, we show 
in Fig.~10 the total isotope temperatures obtained from the same data. 
In addition, we included into this figure also 'total' isotope temperatures 
$T_{H-He}$ obtained in central Au+Au collisions at 50, 100, 150 and 200 
A$\cdot$MeV taken from ref.~\cite{serf}. The corresponding radial flow 
energies were estimated by interpolation of the values given in table~1 
and they are close to other results \cite{Lav,heid,reis3}. 

The found difference between the 'total' and 'thermal' isotope 
temperatures indicates the existence of the two contributions. 
At small flow energies the difference increases with increasing  
flow, because the contribution of the coalescence becomes larger. This 
difference is supposed to decrease at large flow energies, since the 
the contribution of the completely thermalized source disappears. 
At 1~A$\cdot$GeV beam energy, IMF are scarcely produced. Therefore, 
there are large uncertainties in both the radial flow and the isotope 
temperature contributions obtained for the thermal source. However, 
within the error bars no difference between the 'total' and the 
'thermal' chemical temperatures was found. That is consistent with the 
disappearance of the thermalized source. 
The comparison of the 'thermal' and the 'total' temperatures shown in 
Fig.~10 suggests that the transition between 
the described mechanisms accomplishes rather smoothly 
since the temperatures are not very different. 

Figure~10 shows also data obtained from another type of reactions, 
namely break-ups of Au nuclei after peripheral collisions \cite{Schar,hxi}. 
These data correspond to small flow energies 
and they match the general trend indicating 
that the fragment formation seems to correlate with the appearance 
of the radial flow independently of its origin.

From the obtained results we suggest the following evolution of 
the fragment production process with increasing radial flow. 
As long as the flow is small, the fragments are mainly 
produced in a completely equilibrated source. However, the energy available 
for the thermal population of the phase space does not include 
the flow energy. Increasing radial flow influences the 
fragment formation in a twofold way: (i) it increases the velocities 
of nucleons forming a fragment, and (ii) it restricts 
the phase space population by cutting many-particle correlations 
between nucleons. This leads to the 
production of small fragments, which could be effectively described 
as increasing temperature of the thermal source. 
At larger radial flow, only one-particle correlations remain, 
which correspond to the coalescence mechanism. 
Nevertheless, properties of composite particles produced
by coalescence resemble features of statistical processes. 
In particular, one may introduce a {\em common} temperature for
any of the fragments which characterizes their formation probabilities
and which is an ingredient of the chemical equilibrium.  
Therefore, one can formally proceed to treat the fragment formation 
statistically, but taking into account the specific relations 
between the new 'statistical' parameters caused by the dynamics of  
the process (see, e.g., relation (\ref{connec})). Some relations of 
the standard thermodynamics, such as the 
equivalence of the chemical and the kinetic temperatures, are not valid 
in this case. However, one can deduce relations 
between statistical and dynamical parameters, like that shown in Fig.~10. 
\begin{figure}
\begin{center}
\epsfig{file=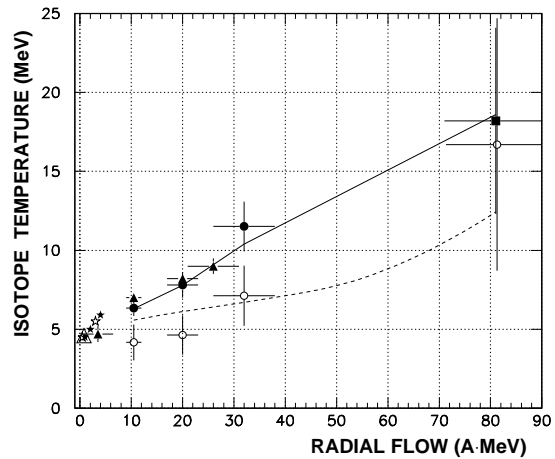,width=8.6cm}
\end{center}
\caption{Isotope temperatures versus radial flow. Open circles: thermal 
temperatures estimated at low fragment velocities.  
Open triangle: thermal temperature taken from \cite{huan}. 
Full symbols correspond to temperatures obtained from
energy-integrated isotope yields.
Dots: FOPI data \cite{pogg}. 
Triangles: ALADIN data \cite{serf}. 
Square: EOS data \cite{eos,lisa}. 
Full asterisks: peripheral collisions from \cite{Schar}. 
Open asterisks: peripheral collisions from \cite{hxi,odeh}. 
Dashed line: Calculated isotope temperature of the thermal source alone. 
Solid line: isotope temperature $T_{H-He}$ calculated by SMMFC including 
             coalescence. 
}
\label{fig:correlation}
\end{figure}

\section{Conclusions} 

We have developed an extended version of the Statistical Multifragmentation Model 
(SMMFC) aimed at the analysis of experimental data related to LCP and IMF 
production in central heavy-ion collisions. In the present studies the 
model was applied to analyse Au+Au data obtained by the FOPI and EOS 
collaborations in the energy range from 100 to 400 and at 
1000 A$\cdot$MeV, respectively. 

The statistical analysis includes the following physical processes  
of the reactions. An essential part of nucleons 
of the whole system is released  during the fast (dynamical) stage 
and the remaining matter constitutes an equilibrium source. The share of 
the complete thermalized source $A_{rel}$ decreases with increasing 
beam energy. This result is in agreement with analyses of experimental 
data carried out in refs. \cite{Lav,will,heid}. 
It was found that the thermal sources have 
temperatures which are considerably lower than expected from the kinetic 
energies of the produced fragments. Both the nucleon and 
the LCP emissions in the fast stage are suppossed to be the 
reason of the reduced equilibrated energy at the freeze-out. Moreover, 
an appreciable part of the available energy of the equilibrated source 
is converted into collective motion, e.g. radial flow. 
Though the analysis remains phenomenologically in part, the found regularities 
seem to be reliable since they are supported also by analyses of 
other data. 
Therefore, the results may be used for interpolation and qualitative 
estimations of parameters of thermal sources. 

According to our findings, a considerable growth of the flow 
energy is accompanied by a very moderate increasing of the thermal
temperature in these central collisions. 
In our opinion, the most reliable determination of the temperature in this 
case should be achieved by the isotope thermometer which can be 
directly related to the thermal source. 
As we have shown, it is possible to apply this thermometer 
for testing the dependence of the temperature on the 
fragment velocities, which can be used to identify 
different mechanisms of fragment production. 

Light clusters can also occur as result of a dynamical process 
which involves secondary interaction of the fast nucleons. 
We have shown that the coalescence mechanism is responsible for the 
production of light fragments and its contribution dominates at larger 
beam energies. Coalescence is caused by the short range attractive 
interaction between the nucleons, and, under some conditions, it 
is consistent with the chemical equilibrium. 
Nevertheless, the isotope temperatures obtained from yields of 
coalescent fragments remain very moderate in comparison with their 
kinetic energies (i.e. with their kinetic temperatures). 
The analysis of experimental LCP energy distributions leads 
also to this conclusion. 
Therefore, one can speculate that the chemical equilibrium 
is attained in such exploding systems. However, 
it is different from the one-particle kinetical equilibration.  
We have demonstrated that the isotope temperature is correlated 
to the radial flow in central collisions. 
This presumed relation between isotope temperature and
radial flow may be used to estimate  
chemical temperatures in different explosive processes.

\vskip3mm 
{\em Acknowledgements} W.N. thanks for a long collaboration within the FOPI project,
A.S.B. thanks the GSI Darmstadt for hospitality and financial support.
The authors thank especially Dr. W. Reisdorf for numerous stimulating 
discussions and Dr. H.W. Barz for valuable comments. We are also indeted to
Drs. H. Prade and F. D\"onau for a careful reading of the manuscript.


\begin{thebibliography}{99}

\bibitem{Bon95}
J.P. Bondorf, A.S. Botvina, A.S. Iljinov, I.N. Mishustin
and K. Sneppen, Phys. Rep.\,{\bf 257}, 133 (1995)

\bibitem{agos}
M. D'Agostino et al., Phys. Letters\,B {\bf 371}, 175 (1996)

\bibitem{Lav}
F. Lavaud, Ph.D. thesis, IPNO T 01-06, IPN Orsay, 2001, unpublished

\bibitem{Bouriq}
B. Boriquet et al., {\em Proceedings of the 39. International
Winter Meeting on Nuclear Physics, Bormio,2001} edited by
I.Iori and A.Moroni (Supplement 117, University of Milano, 2001),
p.\,84--102

\bibitem{will}
C. Williams et al., Phys. Rev.\,C {\bf 55}, R2132 (1997)

\bibitem{neu2}
W. Neubert \,FOPI Collaboration,{\em Proceedings of the International Workshop
on Heavy Ion Physics at Low, Intermediate and High Energies using $4 \pi$ detectors,
Poiana Brasov (Romania), 1996},\,edited by M.Petrovici et al. (World Scientific, 
Singapore 1997),p.\,176

\bibitem{bon2}
J.P. Bondorf et al., Phys. Rev. Lett.\,{\bf 73}, 628 (1994)

\bibitem{Des}
P. Dessesquelles et al.,
Nucl. Phys.\,A {\bf 633}, 547 (1998)

\bibitem{heid}
B. Heide and H.W. Barz, Nucl. Phys.\,A {\bf 588}, 918 (1995)  
and Nucl. Phys.\,A {\bf 591}, 755 (1995)

\bibitem{Schar}
R.P. Scharenberg et al., Phys. Rev.\,C {\bf 64}, 054602 (2001) 

\bibitem{hxi} 
H. Xi et al., Z. Phys.\,A {\bf 359}, 397 (1997)

\bibitem{wallerstein} 
G. Wallerstein et al., Rev. Mod. Phys. {\bf 69}, 995 (1997) 

\bibitem{reis1}
\mbox{W. Reisdorf et al.,\,FOPI collaboration,} Nucl. Phys.\,A 
{\bf 612}, 493 (1997)
and Acta Phys. Polonia\,C {\bf 52}, 443 (1994)

\bibitem{kuhn}
\mbox{C. Kuhn et al.,\,FOPI collaboration,} Phys.Rev.\,C {\bf 48}, 1232 (1993)
 
\bibitem{pogg}
\mbox{G. Poggi et al.,\,FOPI collaboration,}
Nucl. Phys.\,A {\bf 586}, 755 (1993)

\bibitem{eos}
\mbox{M.A. Lisa,\,EOS Collaboration,} {\em Proceedings of the International Workshop
on Heavy Ion Physics at Low, Intermediate and High Energies using $4 \pi$ detectors,
Poiana Brasov (Romania), 1996},\,edited by M.Petrovici et al. (World Scientific, Singapore 1997),
p.\,194 and private communication

\bibitem{lisa}
\mbox{M.A. Lisa et al.\,,EOS collaboration,} Phys. Rev. Lett. {\bf 75}, 2662 (1995)

\bibitem{scott}
A. Scott, Ph.D. thesis, Kent State University, 1995, unpublished
 
\bibitem{siem}
P.J. Siemens and J.O. Rasmussen, Phys. Rev. Lett. {\bf 42}, 880 (1979)

\bibitem{hahn} 
D. Hahn and H. St\"ocker, Nucl. Phys.\,A {\bf 476}, 718 (1988)

\bibitem{fai}
G. Fai and J. Randrup, Comp. Phys. Comm. {\bf42}, 385 (1986)
and Comp.Phys.Comm. {\b77}, 153 (1993)

\bibitem{bot1}
A.S. Botvina et al., Nucl. Phys.\,A {\bf 475}, 663 (1987)

\bibitem{albe} 
S. Albergo et al., Il Nuovo Cimento {\bf 89}, 1 (1985)

\bibitem{moeh}
T. M\"ohlenkamp, Ph.D. thesis, Technische Universit\"at Dresden, 1996,
unpublished

\bibitem{pochodzalla95}
J. Pochodzalla et al., Phys. Rev. Lett. {\bf 75}, 1040 (1995)

\bibitem{kopylov}
G.I. Kopylov, \mbox{{\em Principles of resonance kinematics}},
( Nauka, Moscow,\,1970)

\bibitem{Xi1}
H. Xi et al., Phys. Rev.\,C {\bf 57}, R462 (1998) 

\bibitem{dos1}
K.G.R. Doss et al., Phys. Rev. Lett.\,{\bf 59}, 2720 (1987) and 
Mod. Phys. Lett.\,A{\bf 3}, 849 (1988)

\bibitem{neu1}
W. Neubert and A.S. Botvina, Eur. Phys. J.\,A {\bf 7}, 101 (2000)

\bibitem{BU63}
S.T. Butler and C.A. Person, Phys. Rev. {\bf 129}, 836 (1963)

\bibitem{gutbrod}
H.H. Gutbrod et al., Phys. Rev. Lett. {\bf 37}, 667 (1976)

\bibitem{CS86}
L.P. Csernai and J.I. Kapusta, Phys. Rep. {\bf 131}, 223 (1986)

\bibitem{gross}
D.H.E. Gross, Rep. Progr. Phys. {\bf 53}, 605 (1990)

\bibitem{fermi}
E. Fermi, Prog. Theor. Phys. {\bf 5}, 570 (1950)

\bibitem{mek}
A. Mekjian, Phys. Rev. Lett. {\bf 38}, 640 (1977)

\bibitem{cibor}
J. Cibor et al., Phys. Lett.\,B {\bf 473}, 29 (2000)

\bibitem{bon4}
J.B. Bondorf et al., Nucl. Phys.\,A {\bf 624}, 706 (1997)

\bibitem{lattice}
C.B. Das and S.Das Gupta, Phys. Rev.\,C {\bf 64}, 041601 (2001) and 
F. Gulminelli and Ph. Chomaz, {\bf nucl-th}/0209032, 2002 


\bibitem{petro}
\mbox{M. Petrovici et al.,\,FOPI collaboration,} Phys. Rev. Lett.
{\bf 74}, 5001 (1995)

\bibitem{reis3}
W. Reisdorf and H.G. Ritter\, Annu.Rev.Part.Sci. {\bf 47}, 663 (1997)

\bibitem{stoe}     
H. St\"ocker and W. Greiner, Phys. Reports {\bf 137}, 277 (1986)

\bibitem{aver}
R. Averbeck et al., GSI-Nachrichten 10-95, 1995, unpublished

\bibitem{pelt}
D. Pelte, FOPI collaboration, Z. Phys.\,A {\bf 359}, 55 (1997) and 
Z. Phys.\,A {\bf 357}, 215 (1997)

\bibitem{munz}
C. M\"untz et al., Z. Phys.\,A {\bf 357}, 39 (1997)

\bibitem{bon5}
J.B. Bondorf et al., Phys. Rev.\,C {\bf 58}, R27 (1998)

\bibitem{serf}
V. Serfling et al., ALADIN collaboration, Phys. Rev. Lett. {\bf 80}, 3928 (1998)

\bibitem{odeh}
T. Odeh, Ph.D. thesis 99-15, University of Frankfurt am Main, 1999, unpublished

\bibitem{huan}
M.J.Huang et al., Phys. Rev. Lett. {\bf 78} 1648 (1997)

\end{thebibliography}
\end{document}